\documentclass[twocolumn,letterpaper,aps,prl,superscriptaddress,showpacs,floatfix]{revtex4}

\usepackage{xspace}	
\usepackage{graphicx}	

\newcommand{\dphi}{\Delta \phi}

\newcommand{\pizero}{\mbox{$\pi^{0}$}\xspace}

\newcommand{\beq}{\begin{equation}}
\newcommand{\eeq}{\end{equation}}

\newcommand{\ncoll}{\mbox{$N_{\rm coll}$}\xspace}

\newcommand{\pt}{\mbox{$p_T$}\xspace}

\begin{document}

\title{Suppression of Back-to-Back Hadron Pairs at Forward Rapidity \\
in $d+$Au collisions at $\sqrt{s_{_{NN}}}=200$~GeV }

\newcommand{\abilene}{Abilene Christian University, Abilene, Texas 79699, USA}
\newcommand{\banaras}{Department of Physics, Banaras Hindu University, Varanasi 221005, India}
\newcommand{\barc}{Bhabha Atomic Research Centre, Bombay 400 085, India}
\newcommand{\bnlcoll}{Collider-Accelerator Department, Brookhaven National Laboratory, Upton, New York 11973-5000, USA}
\newcommand{\bnlphys}{Physics Department, Brookhaven National Laboratory, Upton, New York 11973-5000, USA}
\newcommand{\caucr}{University of California - Riverside, Riverside, California 92521, USA}
\newcommand{\charlesczech}{Charles University, Ovocn\'{y} trh 5, Praha 1, 116 36, Prague, Czech Republic}
\newcommand{\chonbuk}{Chonbuk National University, Jeonju, 561-756, Korea}
\newcommand{\ciae}{China Institute of Atomic Energy (CIAE), Beijing, People's Republic of China}
\newcommand{\cns}{Center for Nuclear Study, Graduate School of Science, University of Tokyo, 7-3-1 Hongo, Bunkyo, Tokyo 113-0033, Japan}
\newcommand{\colorado}{University of Colorado, Boulder, Colorado 80309, USA}
\newcommand{\columbia}{Columbia University, New York, New York 10027 and Nevis Laboratories, Irvington, New York 10533, USA}
\newcommand{\czechtech}{Czech Technical University, Zikova 4, 166 36 Prague 6, Czech Republic}
\newcommand{\dapnia}{Dapnia, CEA Saclay, F-91191, Gif-sur-Yvette, France}
\newcommand{\elte}{ELTE, E{\"o}tv{\"o}s Lor{\'a}nd University, H - 1117 Budapest, P{\'a}zm{\'a}ny P. s. 1/A, Hungary}
\newcommand{\ewha}{Ewha Womans University, Seoul 120-750, Korea}
\newcommand{\fit}{Florida Institute of Technology, Melbourne, Florida 32901, USA}
\newcommand{\fsu}{Florida State University, Tallahassee, Florida 32306, USA}
\newcommand{\gsu}{Georgia State University, Atlanta, Georgia 30303, USA}
\newcommand{\hiroshima}{Hiroshima University, Kagamiyama, Higashi-Hiroshima 739-8526, Japan}
\newcommand{\ihepprot}{IHEP Protvino, State Research Center of Russian Federation, Institute for High Energy Physics, Protvino, 142281, Russia}
\newcommand{\illuiuc}{University of Illinois at Urbana-Champaign, Urbana, Illinois 61801, USA}
\newcommand{\inrras}{Institute for Nuclear Research of the Russian Academy of Sciences, prospekt 60-letiya Oktyabrya 7a, Moscow 117312, Russia}
\newcommand{\instpasczech}{Institute of Physics, Academy of Sciences of the Czech Republic, Na Slovance 2, 182 21 Prague 8, Czech Republic}
\newcommand{\isu}{Iowa State University, Ames, Iowa 50011, USA}
\newcommand{\jinrdubna}{Joint Institute for Nuclear Research, 141980 Dubna, Moscow Region, Russia}
\newcommand{\jyvaskyla}{Helsinki Institute of Physics and University of Jyv{\"a}skyl{\"a}, P.O.Box 35, FI-40014 Jyv{\"a}skyl{\"a}, Finland}
\newcommand{\kek}{KEK, High Energy Accelerator Research Organization, Tsukuba, Ibaraki 305-0801, Japan}
\newcommand{\kfki}{KFKI Research Institute for Particle and Nuclear Physics of the Hungarian Academy of Sciences (MTA KFKI RMKI), H-1525 Budapest 114, POBox 49, Budapest, Hungary}
\newcommand{\korea}{Korea University, Seoul, 136-701, Korea}
\newcommand{\kurchatov}{Russian Research Center ``Kurchatov Institute", Moscow, 123098 Russia}
\newcommand{\kyoto}{Kyoto University, Kyoto 606-8502, Japan}
\newcommand{\labllr}{Laboratoire Leprince-Ringuet, Ecole Polytechnique, CNRS-IN2P3, Route de Saclay, F-91128, Palaiseau, France}
\newcommand{\lawllnl}{Lawrence Livermore National Laboratory, Livermore, California 94550, USA}
\newcommand{\losalamos}{Los Alamos National Laboratory, Los Alamos, New Mexico 87545, USA}
\newcommand{\lpc}{LPC, Universit{\'e} Blaise Pascal, CNRS-IN2P3, Clermont-Fd, 63177 Aubiere Cedex, France}
\newcommand{\lund}{Department of Physics, Lund University, Box 118, SE-221 00 Lund, Sweden}
\newcommand{\maryland}{University of Maryland, College Park, Maryland 20742, USA}
\newcommand{\mass}{Department of Physics, University of Massachusetts, Amherst, Massachusetts 01003-9337, USA }
\newcommand{\muenster}{Institut fur Kernphysik, University of Muenster, D-48149 Muenster, Germany}
\newcommand{\muhlenberg}{Muhlenberg College, Allentown, Pennsylvania 18104-5586, USA}
\newcommand{\myongji}{Myongji University, Yongin, Kyonggido 449-728, Korea}
\newcommand{\nagasaki}{Nagasaki Institute of Applied Science, Nagasaki-shi, Nagasaki 851-0193, Japan}
\newcommand{\newmex}{University of New Mexico, Albuquerque, New Mexico 87131, USA }
\newcommand{\nmsu}{New Mexico State University, Las Cruces, New Mexico 88003, USA}
\newcommand{\ornl}{Oak Ridge National Laboratory, Oak Ridge, Tennessee 37831, USA}
\newcommand{\orsay}{IPN-Orsay, Universite Paris Sud, CNRS-IN2P3, BP1, F-91406, Orsay, France}
\newcommand{\peking}{Peking University, Beijing, People's Republic of China}
\newcommand{\pnpi}{PNPI, Petersburg Nuclear Physics Institute, Gatchina, Leningrad region, 188300, Russia}
\newcommand{\riken}{RIKEN Nishina Center for Accelerator-Based Science, Wako, Saitama 351-0198, Japan}
\newcommand{\rikjrbrc}{RIKEN BNL Research Center, Brookhaven National Laboratory, Upton, New York 11973-5000, USA}
\newcommand{\rikkyo}{Physics Department, Rikkyo University, 3-34-1 Nishi-Ikebukuro, Toshima, Tokyo 171-8501, Japan}
\newcommand{\saispbstu}{Saint Petersburg State Polytechnic University, St. Petersburg, 195251 Russia}
\newcommand{\saopaulo}{Universidade de S{\~a}o Paulo, Instituto de F\'{\i}sica, Caixa Postal 66318, S{\~a}o Paulo CEP05315-970, Brazil}
\newcommand{\stonybrkc}{Chemistry Department, Stony Brook University, SUNY, Stony Brook, New York 11794-3400, USA}
\newcommand{\stonycrkp}{Department of Physics and Astronomy, Stony Brook University, SUNY, Stony Brook, New York 11794-3400, USA}
\newcommand{\tenn}{University of Tennessee, Knoxville, Tennessee 37996, USA}
\newcommand{\titech}{Department of Physics, Tokyo Institute of Technology, Oh-okayama, Meguro, Tokyo 152-8551, Japan}
\newcommand{\tsukuba}{Institute of Physics, University of Tsukuba, Tsukuba, Ibaraki 305, Japan}
\newcommand{\vandy}{Vanderbilt University, Nashville, Tennessee 37235, USA}
\newcommand{\waseda}{Advanced Research Institute for Science and Engineering, 
Waseda University, 17 Kikui-cho, Shinjuku-ku, Tokyo 162-0044, Japan}
\newcommand{\weizmann}{Weizmann Institute, Rehovot 76100, Israel}
\newcommand{\yonsei}{Yonsei University, IPAP, Seoul 120-749, Korea}
\affiliation{\abilene}
\affiliation{\banaras}
\affiliation{\barc}
\affiliation{\bnlcoll}
\affiliation{\bnlphys}
\affiliation{\caucr}
\affiliation{\charlesczech}
\affiliation{\chonbuk}
\affiliation{\ciae}
\affiliation{\cns}
\affiliation{\colorado}
\affiliation{\columbia}
\affiliation{\czechtech}
\affiliation{\dapnia}
\affiliation{\elte}
\affiliation{\ewha}
\affiliation{\fit}
\affiliation{\fsu}
\affiliation{\gsu}
\affiliation{\hiroshima}
\affiliation{\ihepprot}
\affiliation{\illuiuc}
\affiliation{\inrras}
\affiliation{\instpasczech}
\affiliation{\isu}
\affiliation{\jinrdubna}
\affiliation{\jyvaskyla}
\affiliation{\kek}
\affiliation{\kfki}
\affiliation{\korea}
\affiliation{\kurchatov}
\affiliation{\kyoto}
\affiliation{\labllr}
\affiliation{\lawllnl}
\affiliation{\losalamos}
\affiliation{\lpc}
\affiliation{\lund}
\affiliation{\maryland}
\affiliation{\mass}
\affiliation{\muenster}
\affiliation{\muhlenberg}
\affiliation{\myongji}
\affiliation{\nagasaki}
\affiliation{\newmex}
\affiliation{\nmsu}
\affiliation{\ornl}
\affiliation{\orsay}
\affiliation{\peking}
\affiliation{\pnpi}
\affiliation{\riken}
\affiliation{\rikjrbrc}
\affiliation{\rikkyo}
\affiliation{\saispbstu}
\affiliation{\saopaulo}
\affiliation{\stonybrkc}
\affiliation{\stonycrkp}
\affiliation{\tenn}
\affiliation{\titech}
\affiliation{\tsukuba}
\affiliation{\vandy}
\affiliation{\waseda}
\affiliation{\weizmann}
\affiliation{\yonsei}
\author{A.~Adare} \affiliation{\colorado}
\author{S.~Afanasiev} \affiliation{\jinrdubna}
\author{C.~Aidala} \affiliation{\mass}
\author{N.N.~Ajitanand} \affiliation{\stonybrkc}
\author{Y.~Akiba} \affiliation{\riken} \affiliation{\rikjrbrc}
\author{H.~Al-Bataineh} \affiliation{\nmsu}
\author{J.~Alexander} \affiliation{\stonybrkc}
\author{A.~Angerami} \affiliation{\columbia}
\author{K.~Aoki} \affiliation{\kyoto} \affiliation{\riken}
\author{N.~Apadula} \affiliation{\stonycrkp}
% ...
\collaboration{PHENIX Collaboration} \noaffiliation

\date{\today}

\begin{abstract}

Back-to-back hadron pair yields in $d+$Au and $p$+$p$ collisions at 
$\sqrt{s_{NN}}=200$~GeV were measured with the PHENIX detector at the 
Relativistic Heavy Ion Collider.  Rapidity separated hadron pairs were 
detected with the trigger hadron at pseudorapidity $|\eta|<0.35$ and the 
associated hadron at forward rapidity (deuteron direction, 
$3.0<{\eta}<3.8$).  Pairs were also detected with both hadrons measured 
at forward rapidity; in this case the yield of back-to-back hadron pairs in 
$d+$Au collisions with small impact parameters is observed to be 
suppressed by a factor of 10 relative to $p$+$p$ collisions. The 
kinematics of these pairs is expected to probe partons in the Au nucleus 
with a low fraction $x$ of the nucleon momenta, where the gluon densities 
rise sharply. The observed suppression as a function of nuclear thickness, 
$p_T$, and $\eta$ points to cold nuclear matter effects arising at high 
parton densities.
 
\end{abstract}

	
\maketitle

Nuclear effects on quark and gluon densities of nucleons bound in nuclei 
can be studied in collisions of deuteron and gold nuclei at the 
Relativistic Heavy Ion Collider.   At $\sqrt{s_{NN}}=200$ GeV, 
the hadron yields in the forward-rapidity 
(deuteron-going) direction were observed to be suppressed for 
$d+$Au collisions relative to $p$+$p$ 
collisions~\cite{rda_brahms,rda_star,rcp_phenix}.  
However, the mechanism for the suppression was not firmly established 
and may indicate novel QCD effects in nuclei.  Competing theoretical 
approaches include initial state energy loss~\cite{vitev2,strikman}, 
parton recombination~\cite{hwa}, shadowing effects~\cite{GSV,vitev}, and 
gluon saturation~\cite{gribov_saturation}.

Back-to-back dijet yields were proposed as an additional observable to 
distinguish better between competing mechanisms. The color glass 
condensate (CGC) framework~\cite{cgc} predicts that quarks and gluons 
scattering at forward angles (large rapidity) will interact coherently off 
gluons at low $x$ in the gold nucleus.  As a result, the rate of observed 
recoiling jets is expected to be suppressed in $d+$Au collisions compared 
to $p$+$p$, and angular broadening of the back-to-back correlation of jets 
is predicted~\cite{monojets,Marquet:2007vb}. Dihadron correlation 
measurements were used~\cite{ida_central_phenix,ida_phenix} successfully in $d+$Au 
collisions to select dijet production based on the back-to-back peak at 
$\Delta\phi$=$\pi$ between trigger hadrons and their associated hadrons. 
Dihadron correlation measurements with varying kinematic constraints 
(transverse momentum \pt and rapidity) probe different $x$ ranges in the 
nucleus. In particular, measurements at forward rapidity are thought to 
probe small $x$ values in the Au nucleus.

In this Letter, we report results on the suppression in $d+$Au relative to 
$p$+$p$ collisions of inclusive \pizero's and back-to-back cluster-\pizero 
pairs in the forward-rapidity region, and for back-to-back \pizero-\pizero 
or hadron-\pizero pairs separated in rapidity. The data were obtained from 
$p$+$p$ and $d+$Au runs in 2008 with the PHENIX detector and include a new 
electromagnetic calorimeter, the muon piston calorimeter (MPC), with an 
acceptance of $3.0<{\eta}<3.8$ in pseudorapidity and $0<{\phi}<2\pi$.  
The clusters are reconstructed from the energy deposit of photons in 
individual MPC towers. 
The MPC comprises 220 PbWO$_4$ towers of 20.2$X_0$ depth, with lateral
dimensions of $2.2$$\times$$2.2$ cm$^2$, and is located $220$ cm along
the beam axis from the nominal interaction point.

The $d+$Au sample is separated into four centrality 
classes -- 0--20\% (most central), 20--40\%, 40--60\%, and 60--88\% (most 
peripheral) -- based on charge deposited in the backward (gold direction) 
beam-beam counter ($3.0<-\eta<3.9$).  We determine the average 
number of binary collisions $\langle\ncoll\rangle$ from a Glauber 
model~\cite{rcp_phenix} and a simulation of the beam-beam counter; $\langle \ncoll 
\rangle$ values are $15.1\pm 1.0$, $10.2\pm 0.70$, $6.6\pm 0.44$, and 
$3.2 \pm 0.19$, respectively.

\begin{figure}[!ht]
  \includegraphics[width=1.0\linewidth]{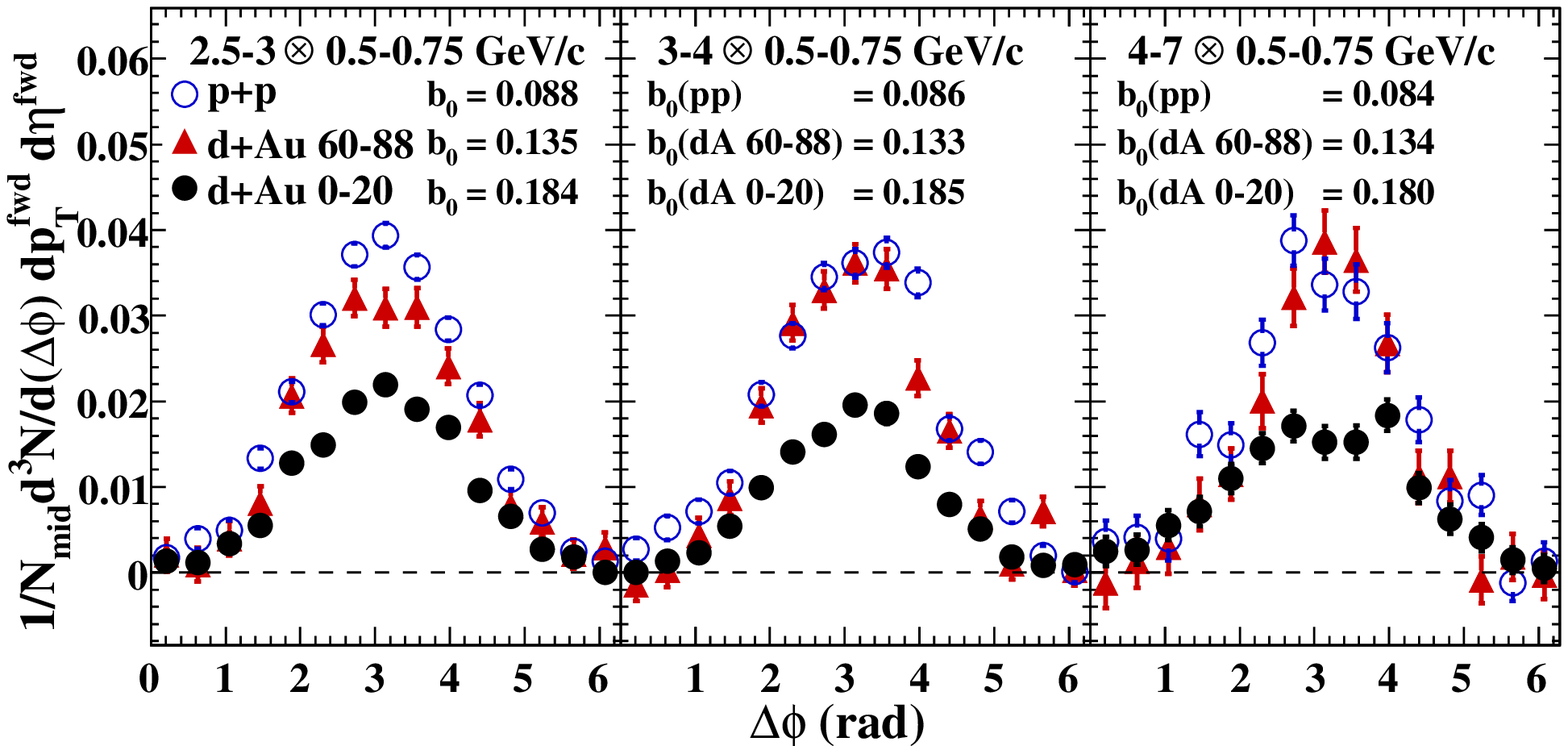}
\caption{(color online).
Pedestal-subtracted \pizero-\pizero per-trigger correlation functions for, 
as indicated, $p$+$p$, $d+$Au peripheral (60--88\% centrality) and $d+$Au 
central (0--20\% centrality) collisions at $\sqrt{s_{NN}}=200$~GeV; the 
associated \pizero's of $p_T=0.5$--0.75 GeV/$c$ are measured at forward 
rapidity $(3.0<\eta<3.8)$ and the triggered \pizero's are measured at midrapidity 
$(|\eta|<0.35)$ for the indicated $p_T$ ranges.  The subtracted pedestal 
values, b$_0$, are also indicated.
}
  \label{fig:fc_corr}

  \includegraphics[width=1.0\linewidth]{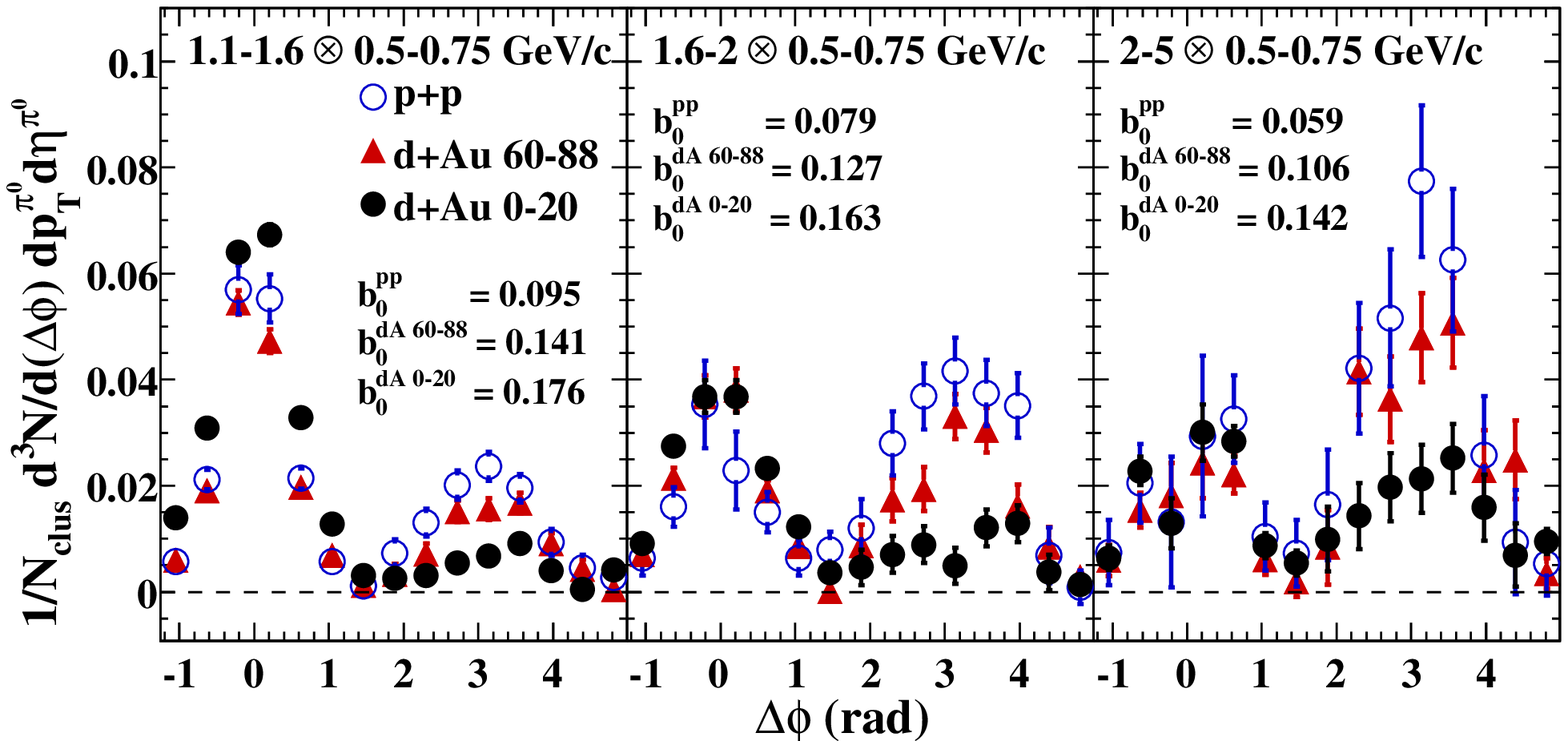}
  \caption{(color online).
Pedestal-subtracted cluster-\pizero per-trigger correlation functions 
measured at forward rapidity $(3.0<\eta<3.8)$ for, as indicated, $p$+$p$, 
$d+$Au peripheral (60--88\% centrality) and $d+$Au central (0-20\% 
centrality) collisions at $\sqrt{s_{NN}}=200$~GeV; the correlation functions are for associated \pizero's 
of $p_T=0.5$--0.75~GeV/$c$ and trigger clusters over the indicated $p_T$ 
ranges. Systematic uncertainties of up to 30\% on the near side 
($|\dphi|<0.5$) are not shown. The subtracted pedestal values, b$_0$, are also indicated.
}
  \label{fig:ff_corr}
\end{figure}

The charged hadron ($h^\pm$) and \pizero analysis in the midrapidity
region $|\eta|<0.35$ is identical to that for previous measurements by
PHENIX~\cite{cntrda,emcrda}.  
For the analysis in the forward-rapidity region,
a fiducial cut is applied ensuring that clusters are
fully reconstructed within the MPC acceptance.
Photon candidates are identified in the MPC by comparing cluster
candidates to the expected shower profile for photons. The shower 
profile for photons is determined from the PHENIX {\sc geant}3~\cite{geant3} based
detector simulation, PISA, which was tuned to reproduce MPC test beam data. 
For the pair analyses reported here, 
the associated particles in the MPC are $\pizero$'s, which are identified by 
reconstructing the mass of their decay photon pairs. The $\pizero$ yield is
obtained after subtraction of the combinatoric background of uncorrelated photon pairs.
The shape of the background was determined from $p$+$p$ {\sc pythia} 6.4~\cite{pythia} and
$d+$Au HIJING~\cite{hijing} events that are subsequently processed through PISA.
The $p_T$-dependent systematic uncertainty on the associated \pizero yield extraction procedure is 
estimated to be 2--5\% for $p$+$p$ and $d+$Au. 

The closeness of the MPC to the collision vertex and the high energy of 
particles emitted in the forward direction make it difficult to 
reconstruct photon pairs from $\pizero$ decays at high $\pt$. For example 
at $\pt=1$~GeV/$c$, approximately 30\% of the photon cluster pairs are merged and cannot 
be reconstructed separately in the MPC. To extend the $p_T$ range and the 
pair yield, single electromagnetic clusters are used as trigger particles 
to construct cluster-$\pi^0$ dihadron pairs in the MPC. 
These trigger clusters are treated assuming that they are all \pizero's.
However, {\sc pythia} studies indicate that $\gtrsim$~80\% of these 
trigger clusters are from \pizero's with the rest being dominantly single photons 
from asymmetric decays of $\eta$ mesons or direct photons;
thus, according to these studies a relatively small contamination remains.
The cluster energy was corrected to the true \pizero energy
to account for the merging effects of the two photons from \pizero decay.
These corrections were determined by embedding Monte Carlo generated
\pizero's into real data, as well as from {\sc pythia} tuned to match the data.

Figure~\ref{fig:fc_corr} shows the azimuthal angle correlations between 
midrapidity and forward-rapidity \pizero pairs, per \pizero trigger 
detected at midrapidity, in $p$+$p$, peripheral $d+$Au, and central $d+$Au 
collisions for varying trigger \pizero $p_T$. Figure~\ref{fig:ff_corr} 
shows the same correlations for trigger clusters where the cluster-\pizero 
pairs are both detected at forward rapidity. The constant pedestal, $b_0$, 
was subtracted from the correlation function. The correlations were 
corrected for the forward \pizero detection efficiency and for the 
combinatoric background beneath the \pizero peaks in the photon-pair 
invariant mass spectra.  This background is determined by measurement of 
the azimuthal correlations for photon-pair mass selections adjacent to the 
\pizero mass window and from studies of simulated jet events from {\sc 
pythia} events processed through PISA.

For the midrapidity/forward-rapidity correlations (Fig.~\ref{fig:fc_corr}), due to the 
large pseudorapidity gap of $\Delta\eta{\sim}3.3$ between the hadrons, 
only an away-side peak ($\Delta\phi$=$\pi$) is seen. For the 
forward-forward correlations a near-side peak ($\Delta\phi$=$0$) is also 
present (see Fig.~\ref{fig:ff_corr}).  The yields and widths of the 
correlated pairs are extracted by fits to an away-side Gaussian signal 
shape plus a constant background ($b_0$). The fit to the forward-forward 
correlations has an additional Gaussian signal for the near-side peak.  
The pedestal is determined from a fit in the midrapidity/forward-rapidity correlations and 
is consistent with the pedestal level found based on the assumption that 
the signal yield is 0 at the minimum of the correlation function - zero 
yield at minimum (ZYAM)~\cite{zyam}.  In the forward-forward correlations 
the ZYAM pedestal is used in the yield extraction. Additional systematic 
uncertainties of up to 30\% (not shown in Fig.~\ref{fig:ff_corr}) are 
ascribed to the near-side peak due to corrections for resonance decays 
that contaminate the jet signal, and due to the acceptance loss around the 
trigger particle of 
$\Delta\phi\times\Delta\eta~\approx~0.5\times0.5$~rad, resulting from 
the minimum separation cut of one tower between cluster peaks in the MPC. 
The acceptance loss gives rise to the decrease observed for the near side peak.

Figures~\ref{fig:fc_corr} and \ref{fig:ff_corr} show that the away-side 
peak for $d+$Au central collisions is suppressed compared to $p$+$p$ 
collisions and peripheral $d+$Au collisions. This effect is large for the 
midrapidity/forward-rapidity correlations (Fig.~\ref{fig:fc_corr}) and becomes even larger 
when both particles are required to be in the forward-rapidity region 
(Fig.~\ref{fig:ff_corr}).

\begin{figure*}[t]
  \includegraphics[width=0.99\linewidth]{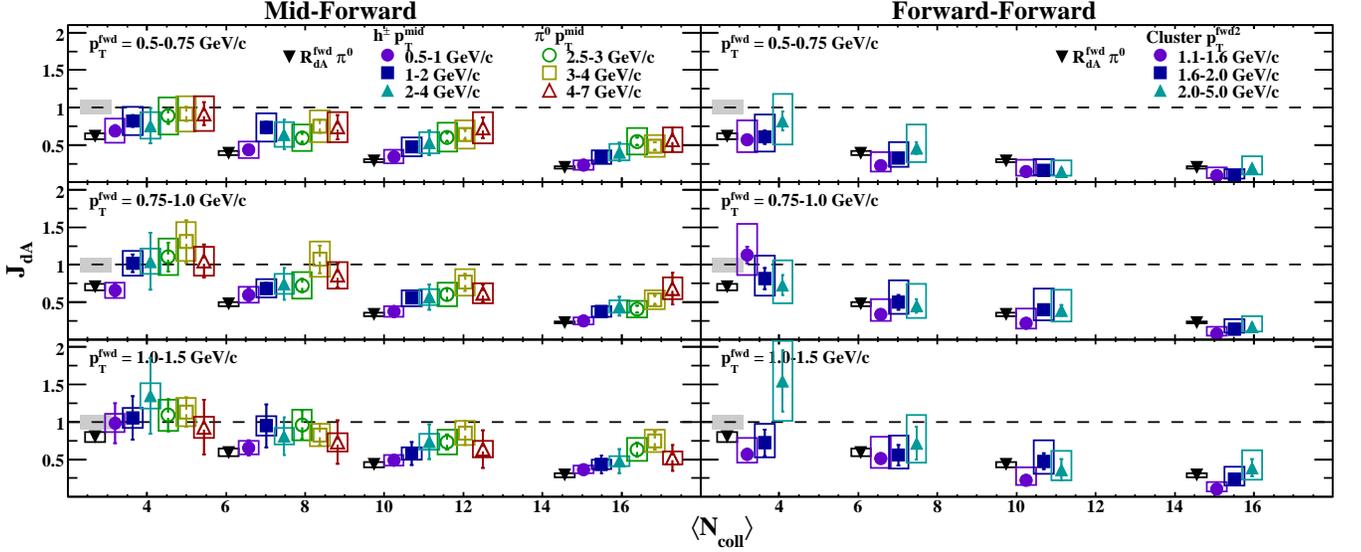}
  \caption{(color online).
Relative yield $J_{dA}$ versus $\langle\ncoll\rangle$ for forward-rapidity 
($3.0<{\eta}<3.8$) \pizero's paired with 
(left) midrapidity ($|\eta|<0.35$) hadrons and \pizero's and 
(right) forward-rapidity ($3.0<{\eta}<3.8$) cluster-\pizero pairs for the 
indicated combinations of $p_T$ ranges.   Also plotted as inverted solid triangles 
are the values of the forward $\pi^0$ $R_{dA}$.  Around each data point the vertical 
bars indicate statistical uncertainties and the open boxes indicate 
point-to-point systematic uncertainties.  The gray bar at the left in each 
panel represents a global systematic scale uncertainty of 9.7\%.  Additional 
centrality dependent systematic uncertainties of 7.5\%, 5.1\%, 4.1\%, and 
4.8\% for the peripheral to central bins, respectively, are not shown.  
The $\langle \ncoll \rangle$ values within a centrality selection 
are offset from their actual values for visual clarity (see text 
for actual $\langle\ncoll\rangle$ values).  }
  \label{fig:jda}
\end{figure*}

For the midrapidity/forward-rapidity correlations, within their large uncertainties the
Gaussian widths of the away-side correlation peak remain the same
between $p$+$p$ and central $d+$Au and the broadening predicted in the
CGC framework in Ref.~\cite{monojets} is not observed.  For
example, in $d$+Au central collisions, 
$\sigma=0.93{\pm}0.09^{\rm stat}{\pm}0.139^{\rm syst}$ 
for $p_T^{\rm fwd}=1.25$ GeV/$c$ and trigger
particle momentum $2.5<p_T^t<3.0$ GeV/$c$, while 
$\sigma=0.97{\pm}0.07^{\rm stat}{\pm}0.08^{\rm syst}$ 
for $p$+$p$ collisions.  For the
forward-forward correlations, the measurement does not discern whether
there is appreciable broadening between $d+$Au and p+p collisions, as
the ZYAM pedestal determination can bias the widths to smaller values.

The observed suppression is quantified by studying the relative yield, 
$J_{dA}$~\cite{jda_orig}, of correlated back-to-back hadron pairs in 
$d+$Au collisions compared to $p$+$p$ collisions scaled with 
$\langle\ncoll\rangle$,
\beq
J_{dA}=I_{dA} \times R^t_{dA}=\frac{1}{\langle\ncoll\rangle}\frac{\sigma_{dA}^{\rm pair}/\sigma_{dA}}{\sigma_{pp}^{\rm pair}/\sigma_{pp}},
\eeq
where $R^t_{dA}=(1/\langle\ncoll\rangle)\cdot(\sigma_{dA}^{t}/\sigma_{dA})/(\sigma_{pp}^{t}/\sigma_{pp})$ is 
the usual nuclear modification factor for trigger particles $t$, and $\sigma$, $\sigma^t$, and 
$\sigma^{\rm pair}$ are the cross sections (or normalized yields) for the full event selection, trigger particle 
event selection, and dihadron pair event selection. $I_{dA}$ is the ratio of conditional hadron yields, $CY$, for
$d+$Au and $p$+$p$ collisions:
\beq 
CY=\frac{\int d(\Delta\phi) [dN/d(\dphi)-b_0]}
          {N^t\times\epsilon^{a}\times\Delta\eta^{a}\times\Delta{p^a_{T}}},
\eeq 
\noindent with the acceptance corrected dihadron correlation function 
$dN/d(\dphi)$, the number of trigger particles $N^t$, the detection 
efficiency for the associated particle $\epsilon^a$ and the level of the 
uncorrelated pedestal in the correlation functions $b_0$. The integral is 
taken over the Gaussian fit of the away-side peak.  The $J_{dA}$ uncertainties  
include a systematic uncertainty from the ZYAM pedestal subtraction. In 
determining this uncertainty it was assumed that changes between $d$+Au 
and $p$+$p$ in the Gaussian away-side width remain below a factor two. 
This upper limit is based on the small observed changes in width in 
the midrapidity/forward-rapidity correlations and the correlations studied previously with 
the PHENIX muon spectrometers~\cite{ida_phenix}.
The $J_{dA}$ is calculated from the measured $I_{dA}$ and $R^t_{dA}$ for 
the forward-rapidity trigger correlations with the new 
$\pi^0$ $R_{d{\rm Au}}=R_{d{\rm Au}}^t$ determined in the MPC. For the 
midrapidity trigger correlations, published values for $R_{dA}$ from the 
2003 RHIC run~\cite{cntrda,emcrda} were used.

\begin{figure}[!ht]
  \includegraphics[width=1.0\linewidth]{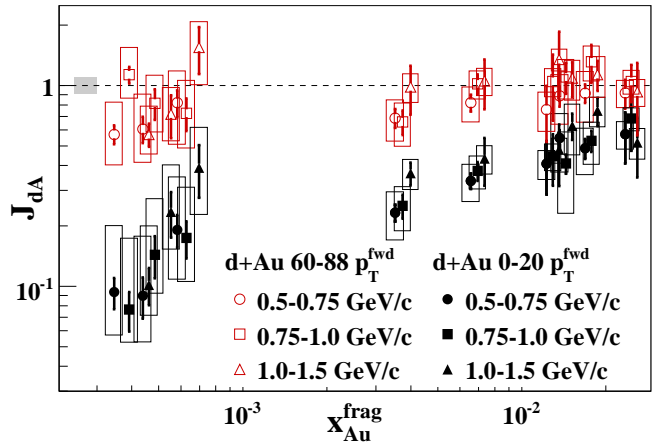}
\caption{(color online).
$J_{dA}$ versus $x_{\rm Au}^{\rm frag}$ for peripheral (60--88\%) and central 
(0--20\%) $d+$Au collisions at $\sqrt{s_{NN}}=200$ GeV.  The 
statistical error bars and systematic uncertainty boxes 
are the same as in Fig.~\ref{fig:jda}.  Above 
$x_{\rm Au}^{\rm frag}>10^{-3}$, some data points were offset from their true 
$x_{\rm Au}^{\rm frag}$ to avoid overlap.  The leftmost point in each 
group of three is at the correct $x_{\rm Au}^{\rm frag}$.}
  \label{fig:jdax}
\end{figure}

Figure~\ref{fig:jda} presents $J_{dA}$ versus $\langle\ncoll\rangle$ for 
forward-rapidity \pizero's paired with midrapidity hadrons and \pizero's, 
and for \pizero's and clusters paired at forward rapidity. The $J_{dA}$ 
decreases with an increasing number of binary collisions, $\langle \ncoll 
\rangle$, or equivalently with increasing nuclear thickness.  The 
suppression also increases with decreasing particle $p_T$ and is 
significantly larger for forward-forward hadron pairs than for midrapidity/forward-rapidity 
pairs. The observed suppression of $J_{dA}$ versus nuclear thickness, 
$p_T$ and $\eta$ points to large cold nuclear matter effects arising at 
high parton densities in the nucleus probed by the deuteron, consistent 
with predictions from CGC~\cite{Marquet:2007vb}.  This trend is 
seen more clearly in Fig.~\ref{fig:jdax} where $J_{dA}$ is plotted versus 
$x_{\rm Au}^{\rm frag}=(\langle p_{T1}\rangle e^{-\langle 
\eta_1\rangle}+\langle p_{T2}\rangle e^{-\langle 
\eta_2\rangle})/\sqrt{s_{NN}}$ for all pair selections in $\eta$ and 
$p_T$.  In the case of $2{\rightarrow}2$ parton scattering, where two 
final state hadrons carry the full parton energy, $z$=$1$, the variable 
$x_{\rm Au}^{\rm frag}$ would be equal to $\langle{x_{\rm Au}}{\rangle}$, 
which is the average momentum fraction of the struck parton in the Au nucleus.  

Because the fragmentation hadrons on average carry a momentum 
fraction $\langle{z}{\rangle}<1$, $x_{\rm Au}^{\rm frag}$ will be smaller 
than $\langle{x_{\rm Au}}{\rangle}$. 
Based on previous studies by PHENIX at
midrapidity, the mean fragmentation $\langle{z}{\rangle}$ is expected
to be between 0.5-0.75~\cite{ppg029}.
In general the theoretical extraction of $x_{\rm 
Au}$ from the measured $p_T$ and $\eta$ will differ from the leading order 
QCD picture of $2{\rightarrow}2$ processes used above.  
Also, at modest 
$p_T$'s the interpretation of the measured correlation functions as high 
energy 2$\rightarrow$2 parton scattering accessing low $x$ may be limited 
by contributions from processes with small momentum transfer $Q^2$.  
Future theoretical analysis will be necessary to evaluate these and other 
contributions from different nuclear 
effects~\cite{vitev2,strikman,hwa,GSV,vitev,gribov_saturation,cgc} on the 
observed large suppression in $J_{dA}$.  These analyses could additionally 
be complicated by the presence of hadron pairs originating from 
multiparton interactions~\cite{strikman_parton} that might not 
probe gluon structure at low $x_{\rm Au}$.

In summary, measurements of the inclusive $\pizero$ yield at forward rapidity, 
of the back-to-back correlated yield of cluster-\pizero pairs
in the forward-rapidity region, and of the correlated yield of 
forward-rapidity $\pizero$'s with midrapidity $\pizero$'s
or hadrons in $p$+$p$ and $d+$Au collisions at $\sqrt{s_{NN}}=200$ GeV 
were presented.  The correlated yields of back-to-back 
pairs were analyzed for various kinematic selections in $\pt$
and rapidity. 
The forward-central pair measurements show no increase
in the azimuthal angular correlation width within experimental uncertainties.
The correlated yield of back-to-back 
pairs in $d+$Au collisions is observed to be substantially suppressed relative 
to $p$+$p$ collisions
with a suppression that is observed to increase with decreasing impact 
parameter selection and for pairs probing more forward rapidities.


We thank the staff of the Collider-Accelerator and Physics
Departments at Brookhaven National Laboratory and the staff of
the other PHENIX participating institutions for their vital
contributions.  We acknowledge support from the 
Office of Nuclear Physics in the
Office of Science of the Department of Energy, the
National Science Foundation, Abilene Christian University
Research Council, Research Foundation of SUNY, and Dean of the
College of Arts and Sciences, Vanderbilt University (U.S.A),
Ministry of Education, Culture, Sports, Science, and Technology
and the Japan Society for the Promotion of Science (Japan),
Conselho Nacional de Desenvolvimento Cient\'{\i}fico e
Tecnol{\'o}gico and Funda\c c{\~a}o de Amparo {\`a} Pesquisa do
Estado de S{\~a}o Paulo (Brazil),
Natural Science Foundation of China (P.~R.~China),
Ministry of Education, Youth and Sports (Czech Republic),
Centre National de la Recherche Scientifique, Commissariat
{\`a} l'{\'E}nergie Atomique, and Institut National de Physique
Nucl{\'e}aire et de Physique des Particules (France),
Ministry of Industry, Science and Tekhnologies,
Bundesministerium f\"ur Bildung und Forschung, Deutscher
Akademischer Austausch Dienst, and Alexander von Humboldt Stiftung (Germany),
Hungarian National Science Fund, OTKA (Hungary), 
Department of Atomic Energy and Department of Science and Technology (India), 
Israel Science Foundation (Israel), 
National Research Foundation and WCU program of the 
Ministry Education Science and Technology (Korea),
Ministry of Education and Science, Russian Academy of Sciences,
Federal Agency of Atomic Energy (Russia),
VR and the Wallenberg Foundation (Sweden), 
the U.S. Civilian Research and Development Foundation for the
Independent States of the Former Soviet Union, 
the US-Hungarian Fulbright Foundation for Educational Exchange,
and the US-Israel Binational Science Foundation.



\begin{thebibliography}{23}
\expandafter\ifx\csname natexlab\endcsname\relax\def\natexlab#1{#1}\fi
\expandafter\ifx\csname bibnamefont\endcsname\relax
  \def\bibnamefont#1{#1}\fi
\expandafter\ifx\csname bibfnamefont\endcsname\relax
  \def\bibfnamefont#1{#1}\fi
\expandafter\ifx\csname citenamefont\endcsname\relax
  \def\citenamefont#1{#1}\fi
\expandafter\ifx\csname url\endcsname\relax
  \def\url#1{\texttt{#1}}\fi
\expandafter\ifx\csname urlprefix\endcsname\relax\def\urlprefix{URL }\fi
\providecommand{\bibinfo}[2]{#2}
\providecommand{\eprint}[2][]{\url{#2}}

\bibitem[{\citenamefont{Arsene et~al.}(2004)}]{rda_brahms}
\bibinfo{author}{\bibfnamefont{I.}~\bibnamefont{Arsene}} \bibnamefont{et~al.}
  (\bibinfo{collaboration}{BRAHMS Collaboration}), \bibinfo{journal}{Phys. Rev.
  Lett.} \textbf{\bibinfo{volume}{93}}, \bibinfo{pages}{242303}
  (\bibinfo{year}{2004}).

\bibitem[{\citenamefont{Adams et~al.}(2006)}]{rda_star}
\bibinfo{author}{\bibfnamefont{J.}~\bibnamefont{Adams}} \bibnamefont{et~al.}
  (\bibinfo{collaboration}{STAR Collaboration}), \bibinfo{journal}{Phys. Rev.
  Lett.} \textbf{\bibinfo{volume}{97}}, \bibinfo{pages}{152302}
  (\bibinfo{year}{2006}).

\bibitem[{\citenamefont{Adler et~al.}(2005)}]{rcp_phenix}
\bibinfo{author}{\bibfnamefont{S.~S.} \bibnamefont{Adler}} \bibnamefont{et~al.}
  (\bibinfo{collaboration}{PHENIX Collaboration}), \bibinfo{journal}{Phys. Rev.
  Lett.} \textbf{\bibinfo{volume}{94}}, \bibinfo{pages}{082302}
  (\bibinfo{year}{2005}).

\bibitem[{\citenamefont{Vitev}(2007)}]{vitev2}
\bibinfo{author}{\bibfnamefont{I.}~\bibnamefont{Vitev}},
  \bibinfo{journal}{Phys. Rev. C} \textbf{\bibinfo{volume}{75}},
  \bibinfo{pages}{064906} (\bibinfo{year}{2007}).

\bibitem[{\citenamefont{Frankfurt and Strikman}(2007)}]{strikman}
\bibinfo{author}{\bibfnamefont{L.}~\bibnamefont{Frankfurt}} \bibnamefont{and}
  \bibinfo{author}{\bibfnamefont{M.}~\bibnamefont{Strikman}},
  \bibinfo{journal}{Phys. Lett.} \textbf{\bibinfo{volume}{B645}},
  \bibinfo{pages}{412} (\bibinfo{year}{2007}).

\bibitem[{\citenamefont{Hwa et~al.}(2005)\citenamefont{Hwa, Yang, and
  Fries}}]{hwa}
\bibinfo{author}{\bibfnamefont{R.~C.} \bibnamefont{Hwa}},
  \bibinfo{author}{\bibfnamefont{C.~B.} \bibnamefont{Yang}}, \bibnamefont{and}
  \bibinfo{author}{\bibfnamefont{R.~J.} \bibnamefont{Fries}},
  \bibinfo{journal}{Phys. Rev. C} \textbf{\bibinfo{volume}{71}},
  \bibinfo{pages}{024902} (\bibinfo{year}{2005}).

\bibitem[{\citenamefont{Guzey et~al.}(2004)\citenamefont{Guzey, Strikman, and
  Vogelsang}}]{GSV}
\bibinfo{author}{\bibfnamefont{V.}~\bibnamefont{Guzey}},
  \bibinfo{author}{\bibfnamefont{M.}~\bibnamefont{Strikman}}, \bibnamefont{and}
  \bibinfo{author}{\bibfnamefont{W.}~\bibnamefont{Vogelsang}},
  \bibinfo{journal}{Phys. Lett.} \textbf{\bibinfo{volume}{B603}},
  \bibinfo{pages}{173} (\bibinfo{year}{2004}).

\bibitem[{\citenamefont{Qiu and Vitev}(2006)}]{vitev}
\bibinfo{author}{\bibfnamefont{J.}~\bibnamefont{Qiu}} \bibnamefont{and}
  \bibinfo{author}{\bibfnamefont{I.}~\bibnamefont{Vitev}},
  \bibinfo{journal}{Phys. Lett.} \textbf{\bibinfo{volume}{B632}},
  \bibinfo{pages}{507} (\bibinfo{year}{2006}).

\bibitem[{\citenamefont{Gribov et~al.}(1983)\citenamefont{Gribov, Levin, and
  Ryskin}}]{gribov_saturation}
\bibinfo{author}{\bibfnamefont{L.}~\bibnamefont{Gribov}},
  \bibinfo{author}{\bibfnamefont{E.}~\bibnamefont{Levin}}, \bibnamefont{and}
  \bibinfo{author}{\bibfnamefont{M.}~\bibnamefont{Ryskin}},
  \bibinfo{journal}{Phys.Rept.} \textbf{\bibinfo{volume}{100}},
  \bibinfo{pages}{1} (\bibinfo{year}{1983}).

\bibitem[{\citenamefont{McLerran and Venugopalan}(1994)}]{cgc}
\bibinfo{author}{\bibfnamefont{L.}~\bibnamefont{McLerran}} \bibnamefont{and}
  \bibinfo{author}{\bibfnamefont{R.}~\bibnamefont{Venugopalan}},
  \bibinfo{journal}{Phys. Rev. D} \textbf{\bibinfo{volume}{49}},
  \bibinfo{pages}{3352} (\bibinfo{year}{1994}).

\bibitem[{\citenamefont{Kharzeev et~al.}(2005)\citenamefont{Kharzeev, Levin,
  and McLerran}}]{monojets}
\bibinfo{author}{\bibfnamefont{D.}~\bibnamefont{Kharzeev}},
  \bibinfo{author}{\bibfnamefont{E.}~\bibnamefont{Levin}}, \bibnamefont{and}
  \bibinfo{author}{\bibfnamefont{L.}~\bibnamefont{McLerran}},
  \bibinfo{journal}{Nucl. Phys.} \textbf{\bibinfo{volume}{A748}},
  \bibinfo{pages}{627} (\bibinfo{year}{2005}).

\bibitem[{\citenamefont{Marquet}(2007)}]{Marquet:2007vb}
\bibinfo{author}{\bibfnamefont{C.}~\bibnamefont{Marquet}},
  \bibinfo{journal}{Nucl. Phys.} \textbf{\bibinfo{volume}{A796}},
  \bibinfo{pages}{41} (\bibinfo{year}{2007}).

\bibitem[{\citenamefont{Adler et~al.}(2006{\natexlab{a}})}]{ida_central_phenix}
\bibinfo{author}{\bibfnamefont{S.~S.} \bibnamefont{Adler}} \bibnamefont{et~al.}
  (\bibinfo{collaboration}{PHENIX Collaboration}), \bibinfo{journal}{Phys. Rev.
  C} \textbf{\bibinfo{volume}{73}}, \bibinfo{pages}{054903}
  (\bibinfo{year}{2006}{\natexlab{a}}).

\bibitem[{\citenamefont{Adler et~al.}(2006{\natexlab{b}})}]{ida_phenix}
\bibinfo{author}{\bibfnamefont{S.~S.} \bibnamefont{Adler}} \bibnamefont{et~al.}
  (\bibinfo{collaboration}{PHENIX Collaboration}), \bibinfo{journal}{Phys. Rev.
  Lett.} \textbf{\bibinfo{volume}{96}}, \bibinfo{pages}{222301}
  (\bibinfo{year}{2006}{\natexlab{b}}).

\bibitem[{\citenamefont{Adler et~al.}(2008)}]{cntrda}
\bibinfo{author}{\bibfnamefont{S.~S.} \bibnamefont{Adler}} \bibnamefont{et~al.}
  (\bibinfo{collaboration}{PHENIX Collaboration}), \bibinfo{journal}{Phys. Rev.
  C} \textbf{\bibinfo{volume}{77}}, \bibinfo{pages}{014905}
  (\bibinfo{year}{2008}).

\bibitem[{\citenamefont{Adler et~al.}(2007)}]{emcrda}
\bibinfo{author}{\bibfnamefont{S.~S.} \bibnamefont{Adler}} \bibnamefont{et~al.}
  (\bibinfo{collaboration}{PHENIX Collaboration}), \bibinfo{journal}{Phys. Rev.
  Lett.} \textbf{\bibinfo{volume}{98}}, \bibinfo{pages}{172302}
  (\bibinfo{year}{2007}).

\bibitem[{\citenamefont{Brun et~al.}(1987)}]{geant3}
\bibinfo{author}{\bibfnamefont{R.}~\bibnamefont{Brun}} \bibnamefont{et~al.},
\bibinfo{note}{{Report No. CERN-DD-EE-84-1}}
  (\bibinfo{year}{1987}).

\bibitem[{\citenamefont{Sjostrand et~al.}(2006)\citenamefont{Sjostrand, Mrenna,
  and Skands}}]{pythia}
\bibinfo{author}{\bibfnamefont{T.}~\bibnamefont{Sjostrand}},
  \bibinfo{author}{\bibfnamefont{S.}~\bibnamefont{Mrenna}}, \bibnamefont{and}
  \bibinfo{author}{\bibfnamefont{P.~Z.} \bibnamefont{Skands}},
  \bibinfo{journal}{JHEP} \textbf{\bibinfo{volume}{0605}}, \bibinfo{pages}{026}
  (\bibinfo{year}{2006}).

\bibitem[{\citenamefont{Wang and Gyulassy}(1991)}]{hijing}
\bibinfo{author}{\bibfnamefont{X.-N.} \bibnamefont{Wang}} \bibnamefont{and}
  \bibinfo{author}{\bibfnamefont{M.}~\bibnamefont{Gyulassy}},
  \bibinfo{journal}{Phys. Rev. D} \textbf{\bibinfo{volume}{44}},
  \bibinfo{pages}{3501} (\bibinfo{year}{1991}).

\bibitem[{\citenamefont{Ajitanand et~al.}(2005)}]{zyam}
\bibinfo{author}{\bibfnamefont{N.~N.} \bibnamefont{Ajitanand}}
  \bibnamefont{et~al.}, \bibinfo{journal}{Phys. Rev. C}
  \textbf{\bibinfo{volume}{72}}, \bibinfo{pages}{011902}
  (\bibinfo{year}{2005}).

\bibitem[{\citenamefont{Adare et~al.}(2008)}]{jda_orig}
\bibinfo{author}{\bibfnamefont{A.}~\bibnamefont{Adare}} \bibnamefont{et~al.}
  (\bibinfo{collaboration}{PHENIX Collaboration}), \bibinfo{journal}{Phys. Rev.
  C} \textbf{\bibinfo{volume}{78}}, \bibinfo{pages}{014901}
  (\bibinfo{year}{2008}).

\bibitem[{\citenamefont{Adler et~al.}(2006{\natexlab{c}})}]{ppg029}
\bibinfo{author}{\bibfnamefont{S.~S.} \bibnamefont{Adler}} \bibnamefont{et~al.}
  (\bibinfo{collaboration}{PHENIX Collaboration}), \bibinfo{journal}{Phys. Rev.
  D} \textbf{\bibinfo{volume}{74}}, \bibinfo{pages}{072002}
  (\bibinfo{year}{2006}{\natexlab{c}}).

\bibitem[{\citenamefont{Strikman and Vogelsang}(2011)}]{strikman_parton}
\bibinfo{author}{\bibfnamefont{M.}~\bibnamefont{Strikman}} \bibnamefont{and}
  \bibinfo{author}{\bibfnamefont{W.}~\bibnamefont{Vogelsang}},
  \bibinfo{journal}{Phys. Rev. D} \textbf{\bibinfo{volume}{83}},
  \bibinfo{pages}{034029} (\bibinfo{year}{2011}).

\end{thebibliography}

\end{document}